\documentclass{iaus}
\usepackage{graphicx}

\title[~~Radio-FIR correlation in nearby galaxies] 
{The Resolved Radio--FIR Correlation in Nearby Galaxies with Herschel and Spitzer}

\author[Tabatabaei et al.]   
{Fatemeh S. Tabatabaei$^1$,
 Eva Schinnerer$^2$,
Eric Murphy$^3 $,
Rainer Beck$^4$,
Annie Hughes$^5$,
Brent Groves$^6$,
\and The KINGFISH Team}

\affiliation{$^1$Max-Planck-Institut f\"ur Astronomie, K\"onigstuhl 17,
69117-Heidelberg, Germany \\ email: {\tt taba@mpia.de} \\[\affilskip]
$^2$Max-Planck-Institut f\"ur Astronomie,  K\"onigstuhl 17,
69117-Heidelberg, Germany \\ email: {\tt schinner@mpia.de} \\[\affilskip]
$^3$The Observatories of the Carnegie Institution for Science, CA 911101, USA \\ email: {\tt emurphy@obs.carnegiescience.edu}\\[\affilskip]
$^4$Max-Planck-Institut f\"ur Radioastronomie,  Auf dem H\"ugel 69,
53121-Bonn, Germany\\ email: {\tt rbeck@mpifr.de} \\[\affilskip]
$^5$Max-Planck-Institut f\"ur Astronomie,  K\"onigstuhl 17,
69117-Heidelberg, Germany \\ email: {\tt hughes@mpia.de} \\[\affilskip]
$^6$Max-Planck-Institut f\"ur Astronomie,  K\"onigstuhl 17,
69117-Heidelberg, Germany \\ email: {\tt brent@mpia.de}}

\pubyear{2011} 
\volume{284}  
\pagerange{1--12}
\jname{The Spectral Energy Distribution of Galaxies}
\editors{R.J. Tuffs \&  C.C.Popescu, eds.}
\begin{document}

\maketitle

\begin{abstract}
We investigate the correlation between the far-infrared (FIR) and  radio continuum emission from NGC6946 on spatial scales between 0.9 and 17\,kpc. We use the Herschel PACS (70, 100, 160$\mu$m) and SPIRE (250$\mu$m) data from the KINGFISH project. Separating the free-free and synchrotron components of the radio continuum emission, we find that FIR is better correlated with the free-free than the synchrotron emission. Compared to a similar study in M33 and M31, we find that the scale dependence of the synchrotron--FIR correlation in NGC6946 is more similar to M31 than M33. 
The scale dependence of the synchrotron--FIR correlation can be explained by the turbulent-to-ordered magnetic field ratio or, equivalently, the diffusion length of the cosmic ray electrons in these galaxies.
\keywords{Star formation, ISM, Magnetic field, Cosmic ray electrons.}
\end{abstract}

\firstsection 
\section{Introduction}
The correlation between the radio and far-infrared (FIR) luminosities of galaxies
has been shown to be invariant over more than 4 orders of magnitude and out to a
redshift of $z~\sim~3$ (e.g. Sargent et al. 2010). It is only recently that variations in the radio-FIR
correlation have become apparent when studying correlations on different scales
within galaxies. Detailed multi-scale analysis of the radio and FIR 
 data showed that the smallest scale on which the radio-FIR
correlation holds is not the same in the LMC (Hughes et al. 2006), M51 (Dumas et al. 2010), M31 and M33 (Tabatabaei et al. 2007a, submitted); the reason is still unclear. Could massive star formation alone 
cause such variations? What is the influence of the dust heating sources? Are differences in the magnetic 
field structure important? And what is the role of the propagation of cosmic ray electrons (CREs)?

One fact that makes the link between star formation and FIR less obvious is that FIR 
emission consists of at least two emission components: cold dust and warm dust. 
The cold dust emission may not be directly linked to the young stellar population, 
but is rather powered by the interstellar radiation field (ISRF, Xu 1990). Furthermore, it is 
necessary to distinguish between the two main radio continuum (RC) components, 
free-free and synchrotron emission.  The free-free  emission from electrons 
in HII regions around young, massive stars is expected to be closely connected to 
the warm dust emission that is heated by those same stars. Although the synchrotron-emitting CREs also originate  from star forming regions  (i.e. supernovae remnants; the final episodes 
of massive stars), the synchrotron--FIR correlation may not be as tight as the 
free-free--FIR correlation {\it locally}, as a result of diffusion of the cosmic ray 
electrons from their production sites. 


\section{Free-free and synchrotron emission from NGC\,6946}
Tabatabaei et al. (2007b) presented a method to separate the free-free and synchrotron components of the RC emission from M\,33, in which a de-reddened H$\alpha$ map was used as a template for the free-free emission.  Using the same method, we derive the free-free  emission from NGC6946 at 20\,cm. Subtraction of the free-free emission from the observed  20\,cm (VLA + Effelsberg, Beck 1991) emission then yields the synchrotron map (Fig.\,1). Diffusion/propagation of the CREs is evident from the synchrotron map, which shows a diffuse extended emission as well as strong emission from the galaxy center, giant star forming regions, and spiral arms.  The thermal free-free emission has a more clumpy distribution following star forming regions. The thermal fraction is about 8\% at 20\,cm.

\begin{figure}
\begin{center}
\resizebox{\hsize}{!}{\includegraphics*{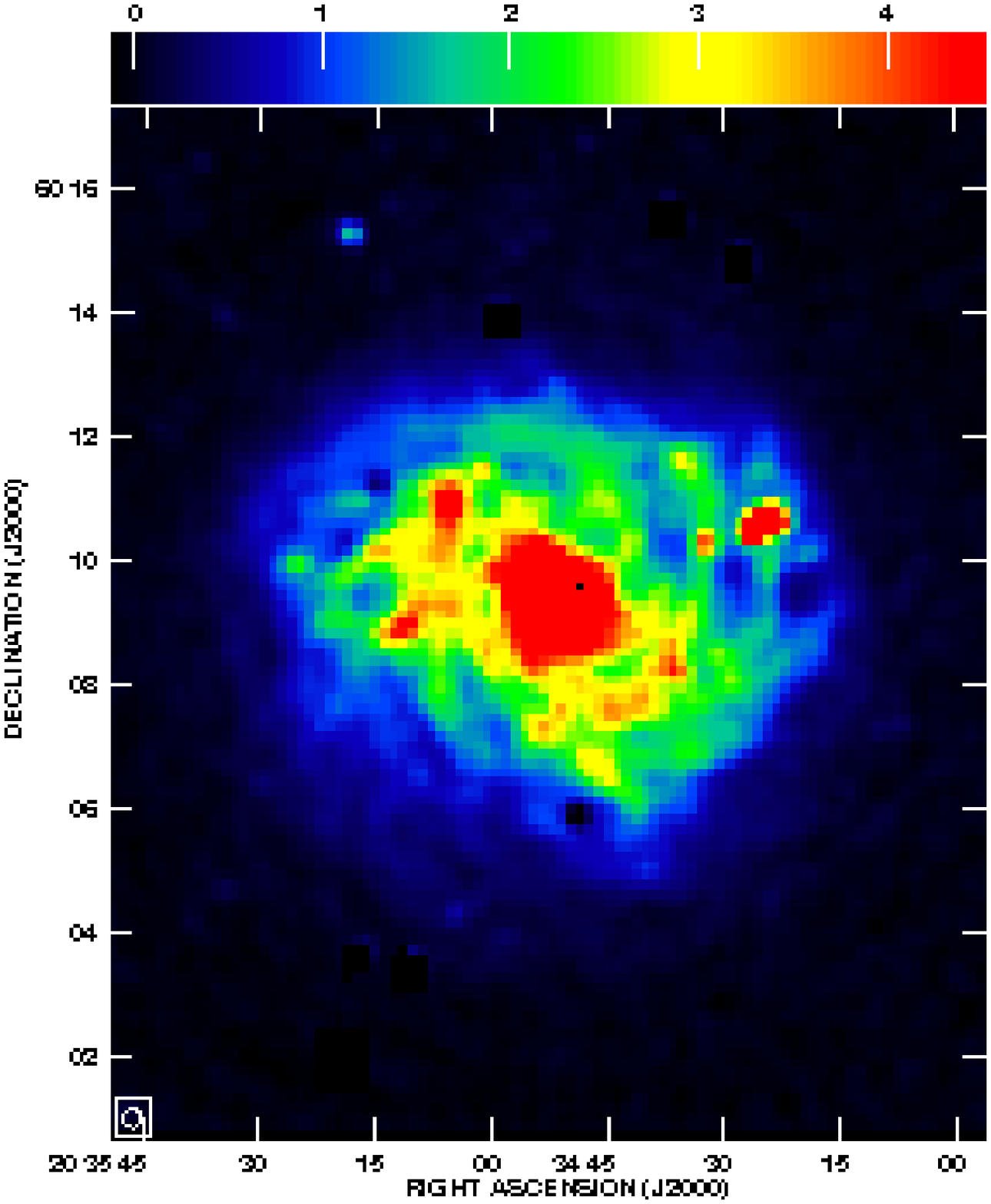}\includegraphics*{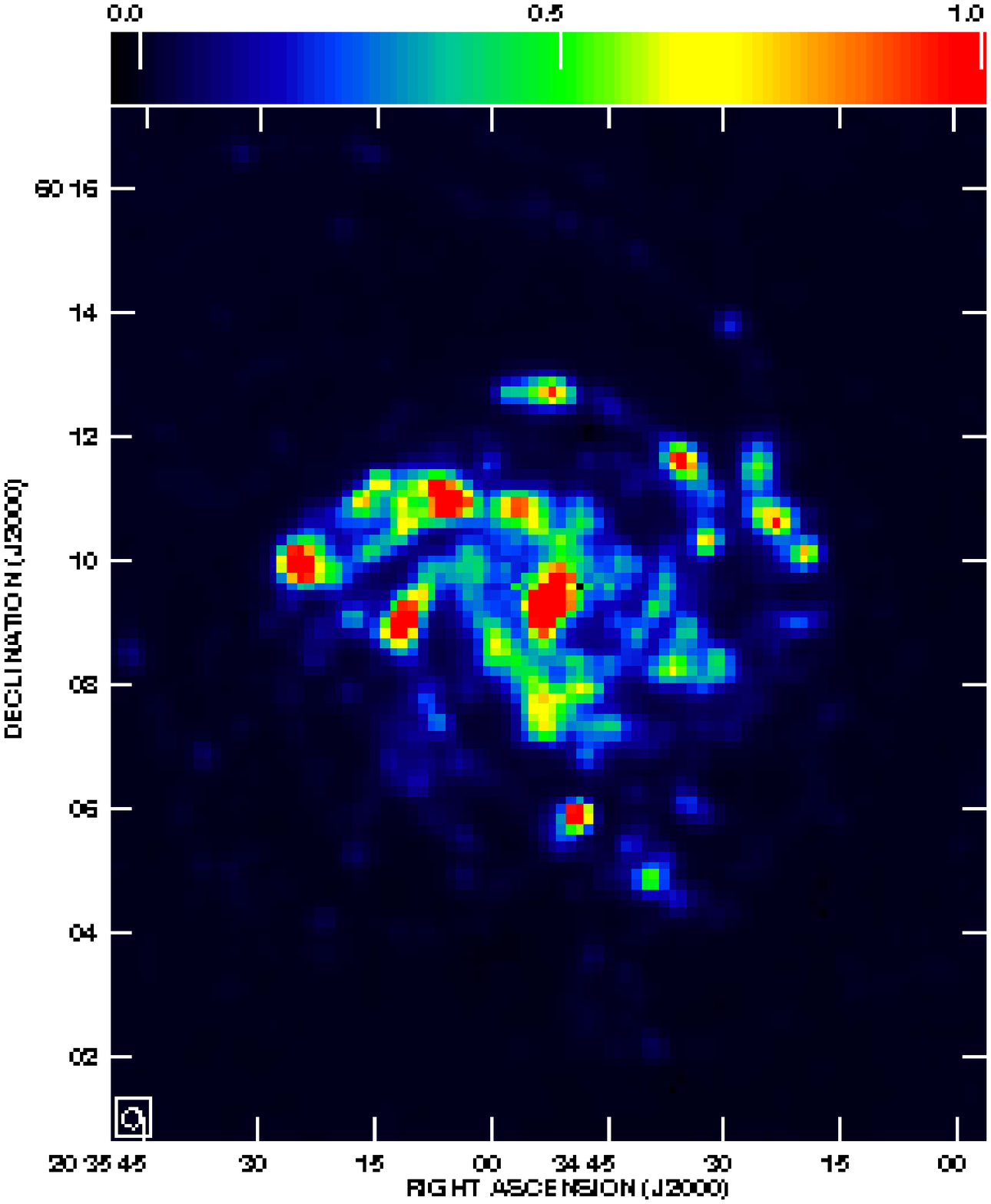}}
\caption[]{20cm synchrotron (left) and free-free emission (right) from NGC6946 at 18'' resolution.  The units are in mJy/beam.  Here external radio sources have been blanked. }
\end{center}
\end{figure}

\section{Wavelet correlation}
Using a 2D continuous wavelet transformation ('Pet-hat', Frick et al. 2001), we decompose the Herschel PACS/SPIRE FIR and the  20cm maps into 8 spatial scales ({\it a}) from 0.9 ($\sim$ twice our linear resolution of 18'') to 17\,kpc. For each scale of decomposition, we derive cross-correlation coefficients  ($r_w$) between the FIR and RC maps. Figure 2 shows $r_w(a)$  between the 70$\mu$m and the free-free/synchrotron emission in NGC\,6946, compared with similar correlations in M\,33 and M\,31 (Tabatabaei et al. submitted).  
\begin{figure}
\begin{center}
\resizebox{\hsize}{!}{ \includegraphics*{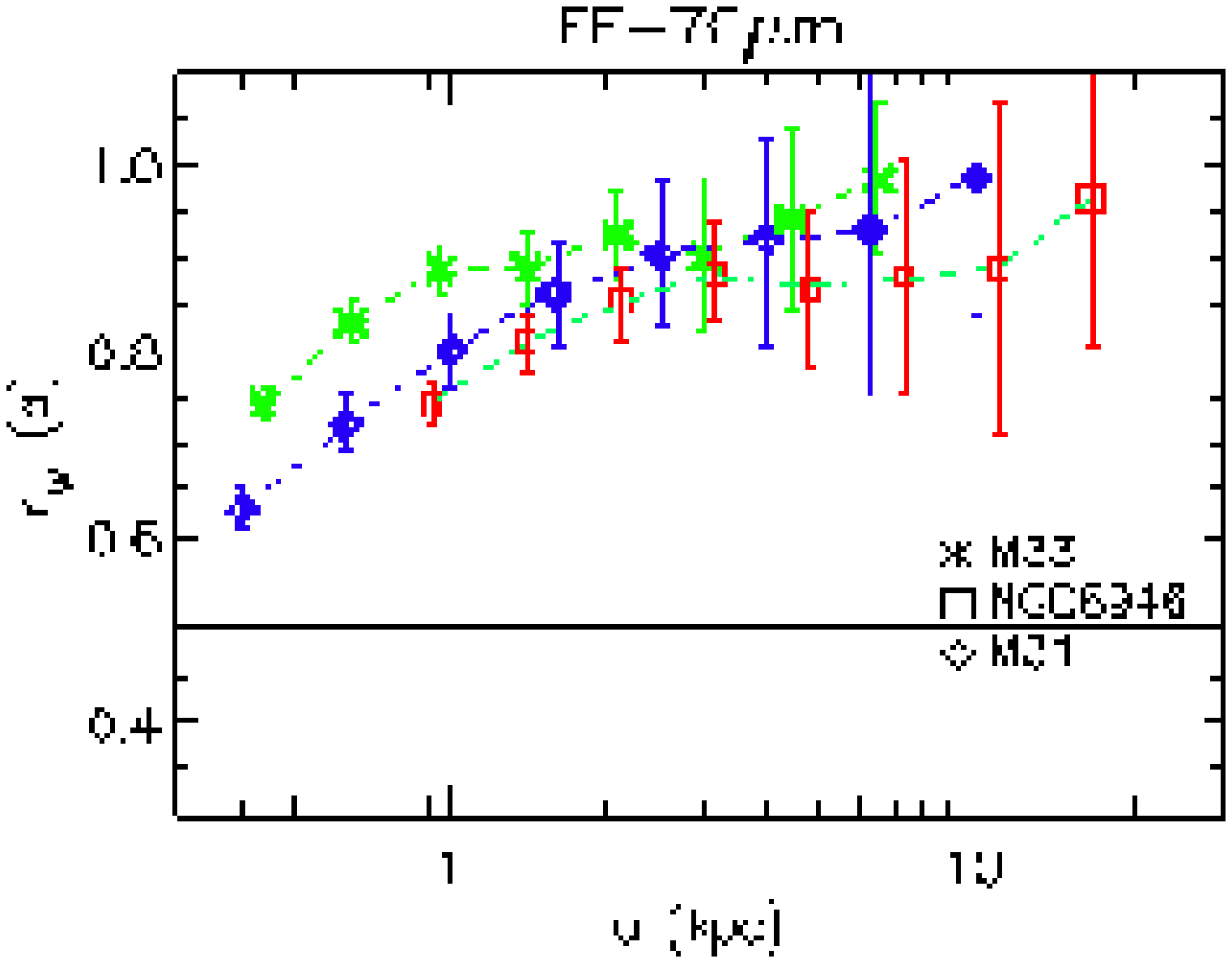}\includegraphics*{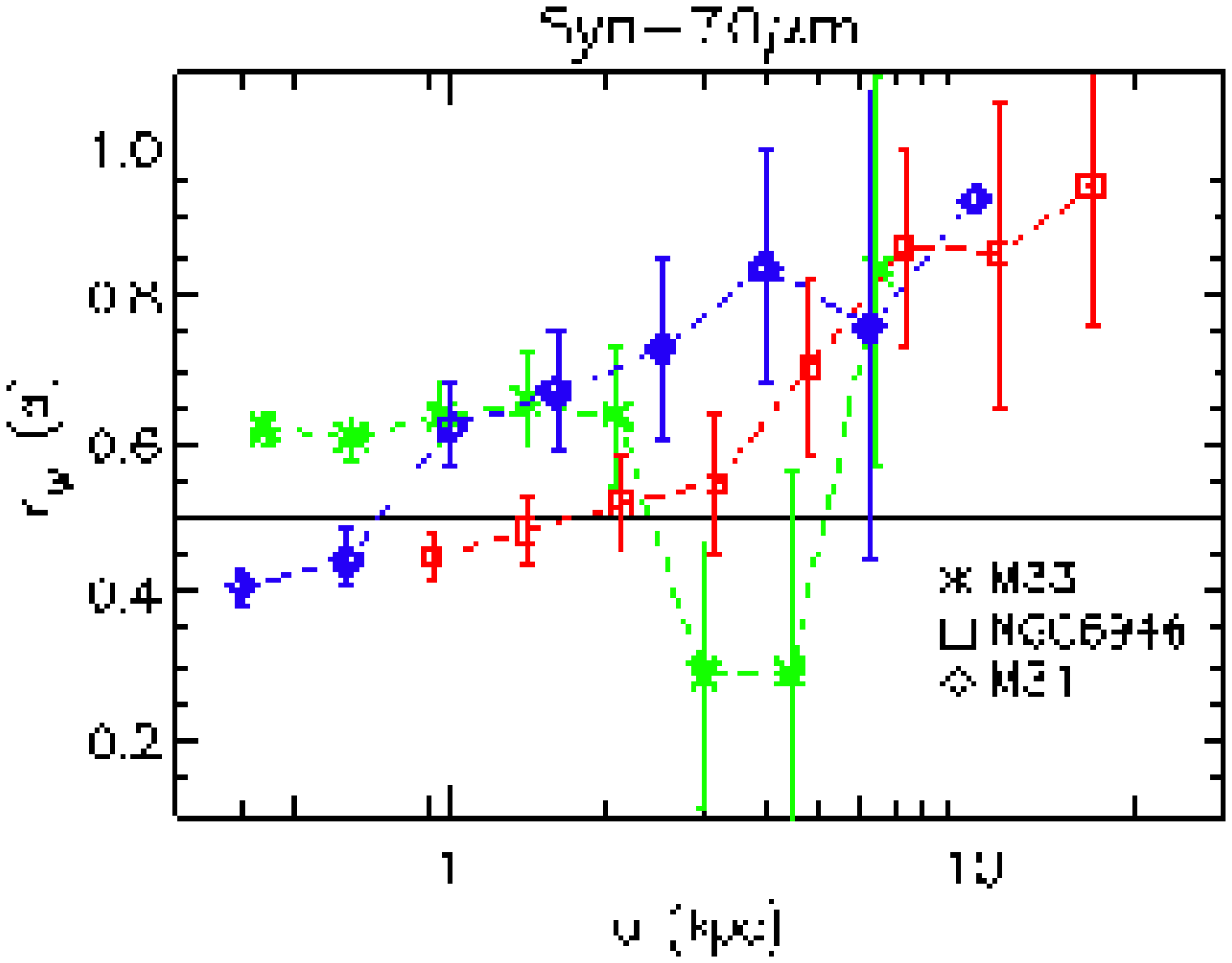}}
\caption[]{Multi-scale correlations between 70$\mu$m FIR and free-free (left), synchrotron (right) emission from NGC6946, compared to those in M31 and M33. }
\end{center}
\end{figure}

\section{Results}
The free-free--FIR correlation is stronger than the synchrotron--FIR correlation in NGC\,6946 as seen in the other two galaxies.  Interestingly, the free-free--FIR correlation exhibits a more or less similar trend in these galaxies, unlike the synchrotron--FIR correlation which shows more variation with scale $a$.  In NGC\,6946, the synchrotron--FIR correlation decreases towards  smaller scales. Such a trend is also seen in M31, M51 (Dumas et al. 2011), and the LMC (Hughes et al. 2006) but not in M33 which shows a stronger correlation on small scales and weaker on larger scales (3-5\,kpc).   
What causes these variations? Assuming that massive stars are the common source of the free-free, FIR (via heating the dust), and synchrotron emission, the synchrotron--FIR variations must be linked to variations in magnetic fields (structure and strength) and diffusion/propagation of CREs.  One would expect a better correlation on small scales ($\sim$ 100\,pc) in galaxies with higher star formation rate (SFR) due to the stronger turbulent magnetic field and/or high concentration of young CREs near their star forming regions (e.g. Murphy et al. 2006).
This is evident for M33 which has a 5 times higher  SFR surface density, $\Sigma ({\rm SFR})$, than M31 (Tabatabaei \& Berkhuijsen 2010). This cannot be investigated in NGC6946, since there is no information for scales $<$ 1\,kpc. On the other hand, 
including the LMC and M51 suggests that the smallest scale on which the radio--FIR correlation holds ($r_w>0.5$),  is better correlated with the ratio of turbulent-to-ordered  magnetic field, which in turn is related to the diffusion length of CREs (Yan \& Lazarian 2004), rather than with $\Sigma ({\rm SFR})$ (Tabatabaei et al. submitted).   

On large scales ($a>$ 1\,kpc),  the synchrotron emission is due to an ordered (uniform) magnetic field and diffused, older CREs.  The ordered magnetic field is controlled by various dynamical, environmental effects caused by, e.g., galactic rotation, density waves, shear and anisotropic compressions in spiral arms (e.g. Fletcher et al. 2011). A good synchrotron--FIR correlation on large scales is expected when enough diffuse CREs exist and that they have opportunity to explore the gaseous/dusty components of a galaxy (Hughes et al. 2006). This  would require a confinement of the magnetic field to the gas. We plot a schematic view of this condition in Fig.\,3, showing sources of the FIR and synchrotron emission and factors which control their correlation on small and large scales.

\begin{figure}
\begin{center}
\resizebox{9cm}{!}{\includegraphics*{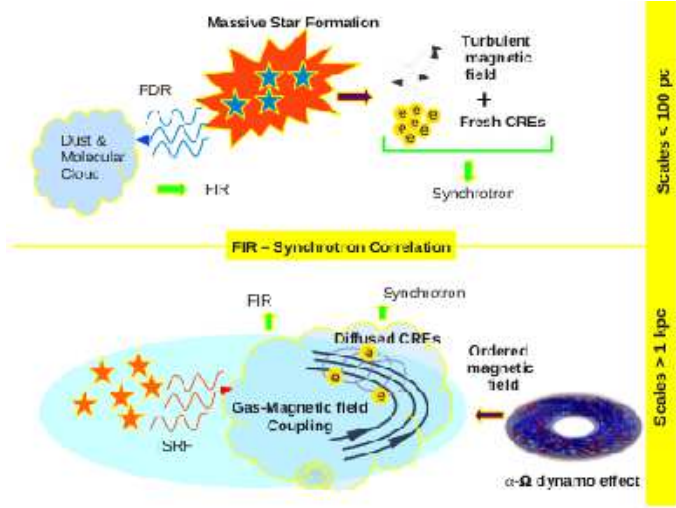}}
\caption[]{A sketch of various sources of the synchrotron and FIR emission and important factors controlling the synchrotron-FIR correlation on small and large scales in a galaxy.}
\end{center}
\end{figure}

\section{Summary}

We derived radio--FIR correlations as a function of scale for 
NGC6946, separately for free-free and synchrotron emission, and for various Herschel FIR bands (70, 100, 160, and 250$\mu$m). Our main results are summarized as follows:\\  
- We find strong synchrotron emission from the central part and around star-forming regions. This is explained by the direct proportionality between SFR and turbulent magnetic field/ young CREs. In addition a smooth diffuse emission is present.\\
- The wavelet power spectrum of the synchrotron emission is stronger on scales $>1$\,kpc than $<1$\,kpc, possibly due to fast diffusion of CREs in the strong, ordered magnetic field.\\
- The  FIR and free-free emissions are correlated on all scales; a better correlation is found with warmer dust.\\
- The FIR and synchrotron emission may not be correlated on some scales, due to diffusion of CREs and/or magnetic field structures. \\
Last but not least, our results show that the radio-FIR correlation depends not only on star formation but also on properties of the magnetic field in galaxies.


\begin{thebibliography}{}

\bibitem[Beck (1991)]{Beck}
{Beck, R.} 1991,
\textit{A}\&\textit{A}, 251, 15

\bibitem[Dumas et al. (2011)]{Dumas}
{Dumas, G., Schinnerer, E., Tabatabaei, F. S., et al.} 2011,
\textit{AJ}, 141, 41

\bibitem[Fletcher et al. (2011)]{Fletcher_11}
{Fletcher, A., Beck, R., Shukurov, A., Berkhuijsen, E. M., \& Horellou, C.} 2011,
\textit{MNRAS}, 412, 2396

\bibitem[Frick et al. (2001)]{Frick_etal_01}
{Frick, P., Beck, R., Berkhuijsen, E. M., \& Patrickeyev, I.} 2001,
\textit{MNRAS}, 327, 1145

\bibitem[Hughes et al. (2006)]{Hughes_etal_06}
{Hughes, A., Wong, T., Ekers, R., et al. } 2006, 
\textit{MNRAS}, 370, 363

\bibitem[Murphy et al. (2006)]{Murphy_06}
{Murphy, E. J., Helou, G., Braun, R., et al. } 2006,
\textit{ApJL}, 651, L111

\bibitem[Tabatabaei et al. (2007a)]{Tabatabaei_1_07} 
{Tabatabaei, F. S., Beck, R., Krause, M., et al. } 2007b,
\textit{A\&A}, 466, 509

\bibitem[Tabatabaei et al. (2007b)]{Tabatabaei_2_07} 
{Tabatabaei, F. S., Beck, R., Kr\"ugel, E., et al. } 2007a,
\textit{A\&A}, 475, 133

\bibitem[Tabatabaei \& Berkhuijsen (2010)]{Tabatabaei_10}
{Tabatabaei, F. S. \& Berkhuijsen, E. M.} 2010,
\textit{A\&A}, 517, A77

\bibitem[Xu (1990)]{Xu_90}
{Xu, C.} 1990,
\textit{ApJ}, 365, 47

\bibitem[Yan \& Lazarian (2004)]{Yan}
{Yan, H., Lazarian, A.} 2004, 
\textit{ApJ}, 614, 757


\end{thebibliography}
\end{document}